\documentclass[3p,times]{elsarticle}

\usepackage{ecrc}
\usepackage{multirow}
\usepackage[bookmarksnumbered, pdfstartview=FitH,colorlinks,urlcolor=blue, citecolor=blue,linkcolor=blue,] {hyperref}
\usepackage{upgreek}

\volume{00}

\firstpage{1}

\journalname{Nuclear Instruments and Methods in Physics Research Section A}

\runauth{H. Miao, W. Chen}

\jid{procs}

\jnltitlelogo{}

\usepackage{amssymb}
\usepackage{amsmath}
\usepackage{subfigure}
\usepackage{lineno}

\usepackage[figuresright]{rotating}
\usepackage{setspace}
\usepackage[bookmarksnumbered, pdfstartview=FitH,colorlinks,urlcolor=blue, citecolor=blue,linkcolor=blue,] {hyperref}

\biboptions{sort&compress}

\begin{document}

\doublespacing


\begin{frontmatter}




\title{MCPSim: A Geant4-based generic simulation toolkit for electron multipliers represented by Microchannel Plate}


	\author[a,b]{~Han Miao}
	\author[a,c]{~Huaxing Peng\footnote{Email: penghuaxing@ihep.ac.cn}}
	\author[a,c]{~Baojun Yan\footnote{Email: yanbj@ihep.ac.cn}}
	\author[a,b,c]{~Shulin Liu}
	\author[a,b]{~Hai-Bo Li}

	\address[a]{\it Institute of High Energy Physics, Beijing 100049, People's Republic of China}
	\address[b]{\it University of Chinese Academy of Sciences, Beijing 100049, People's Republic of China}
	\address[c]{\it State Key Laboratory of Particle Detection and Electronics, Beijing, 100049, People's Republic of China}

\begin{abstract}
\doublespacing

	The simulation of the instruments based on the cascade multiplication of electrons has always been an important and challenging subject in no matter high energy physics, astrophysics, radiography and other fields. In this work, a generic simulation toolkit is developed based on Geant4, ROOT and CADMesh toolkits for the electron multipliers based on the emission process of secondary electrons, especially for Microchannel Plate (MCP). Geometry of instruments can be directly imported by standard CAD files and multiple kinds of electromagnetic fields are supported. 
MCPSim is capable for simulating the general hadronic, weak and electromagnetic processes together with secondary electron emission, which provides a wider range of applications. Users are able to configure the simulation and define the output information by a simple textfile. By comparing with experimental measurements, good agreement is found. MCPSim is expected to guide the design and optimization of such electron multipliers and will be continuously developed and maintained to better satisfy the present and future requirements.

\end{abstract}

\begin{keyword}
Simulation \sep Electron multiplier \sep MCP \sep Geant4



\end{keyword}

\end{frontmatter}


\section{Introduction}

In the past several decades, electron multipliers based on secondary electron emission process, especially the Microchannel Plate (MCP) have been widely used in high energy physics~\cite{PANDACherenkovGroup:2019bmc, Belle-II:2010dht}, astrophysics~\cite{X-ray-camera:1986, astrophysics:2007, X-ray:1994}, radiography~\cite{LEHMANN2004228, Preston_2011}, medical imaging~\cite{Williams1998, Choong2010, Heejong2009, Kim2010} and many other fields due to its excellent spatial and time resolution. To better instruct the design and optimization of such electron multipliers, simulation is essential but has always been suffering from the description of secondary electron emission process. From the former researches, the secondary electrons can be divided to three kinds: backscattering, rediffused and true secondary electrons~\cite{Dekker1958}. To describe the secondary electron emission process quantitatively, models are developed from both theoretical and phenomenological point of view. In 1973, Gerald~F.~Dionne developed a model describing the relationship between the production yield of true secondary electrons and the energy of incident electron based on several reasonable assumption~\cite{Dionne1973, Dionne1975}, which is usually used for analyzing the characterastics of secondary electron emission for different materials~\cite{Guo2019, Grais1982, Kazemian2006}. Later, the model is further proceeded by David~C.~Joy~\cite{Lin2005}. At the same period, a model is raised by M.~S.~Chung and T.~E.~Everhart in 1974, providing the spectrum of the escaping true secondary eletrons from metal surface~\cite{Chung1974}. In 2002, M.~A.~Furman demonstrated a phenomenological model based on the experimental measurements and the method of probability as well as statistics rather than the fundamental physics processes~\cite{Furman2002}. Furman model is widely used in the Monte Carlo simulation of the secondary electron emission process and varified by multiple works~\cite{Ozok2017, Wu2008, Cheng2009, Xie2018}, which is also used in MCPSim and will be introduced for detail in Section~\ref{sec:Furman}. Except from the models above, there are other computional models developed in history~\cite{Shymanska2010, Shymanska2014, Price2001}. However, we still have to admit that the secondary electron emission process is really complex and hard to be modelled due to the large uncertainty of the dynamics of particles and the fluctuation of statistics.

Based on the various models, much effort has been put into the simulation of the electron multiplier devices all over the world~\cite{Ozok2017, Wu2008, Cheng2009, Xie2018, Wang2016, Kruschwitz2008, Guo2021, Chen2019, Li2020, Cai2010, Han2013, Hideki0000, Chen2017, Ma2018, Ma2018-2, Craig2011, Zhao2020, Creusot2010}. Several commercial softwares are developed for the general simulation of electron multipliers including CST Studio $\rm Suite^{TM}$~\cite{CST}, $\rm SIMION^{TM}$~\cite{SIMION} and so on. However, an open-source simulation toolkit will still benefit the communities such as high energy physics and other fields, which is the original aim for us to develop MCPSim. In view of the possible applications, Geant4~\cite{GEANT4:2002zbu, Allison:2006ve, Allison:2016lfl} and ROOT~\cite{ROOT} is chosen as the foundations of MCPSim. There have been works about simulation secondary electron emission using Geant4. The response of multiplication devices in such simulations usually depends on experimental measurements~\cite{Kim2010} and models or packages external from Geant4~\cite{Ma2018}, while there are still helpful attempts to simulate the process track by track~\cite{Ozok2017} although it is meant to a heavy burden for calculation. We also met this problem during the development of MCPSim, we have to admit that it is still a severe challenge for us although we have try to make the simulation quicker in multiple aspects. In 2021, the secondary electron emission process is implemented to Geant4 toolkit~\cite{Gibaru2021} based on the fundamental theories of condensed matter physics and solid state physics and is tested to be agree well with experimental results for some specific materials like C, Al, Si and so on. Since the implementation is finished after the main work of MCPSim, it is not complemented to MCPSim but will be a possible supplement in future. Considering that Furman model~\cite{Furman2002} is used in most simulation works and can be applied to abundant situations as long as valid parameters are accessible, it is implemented to MCPSim and compared with experimental results. Except the secondary electron emission process, other hadronic, weak and electromagnetic processes are simulated by the official Geant4 physics list so that MCPSim can be applied for a wider range. The default physics list is a customized one specially adjusted for the simulation of electrons with low energy. Users are allowed to change the default physics list by a simple parameter in MCPSim.

The article is organized in the following structure. Section~\ref{sec:Furman} will introduce simply the model of secondary electron emission and how we implement it into MCPSim. Section~\ref{sec:geo_and_field} is how to configure the simulation using MCPSim. 
Section~\ref{sec:event_model} will briefly explain the output data structure. The comparison with experimental results are shown in Section~\ref{sec:comparison}. The last part, Section~\ref{sec:summary} will be the summary and some discussion about the possible improvements in future.

\section{Implementation of secondary electron emission process in Geant4}
\label{sec:Furman}

For a given incident electron, the parameters of secondary electrons are got by sampling the secondary electron yield (SEY) distribution, angular distribution and spectrum. The average SEY (aSEY) when the incident electron hit the surface with an angle $\theta$ with respect to the normal direction can be described by~\cite{Scholtz1997, Shih1993, Salehi1981, Kawata1992, Suharyanto2007}:
\begin{equation}
	\label{equ:average_SEY}
	{\rm aSEY}(\theta) = {\rm aSEY}(0) \cdot e^{-\frac{L}{\lambda}(1-{\rm cos}\theta)},
\end{equation}
where $L$ is the average track length inside the material, which is a constant for given material, and $\lambda$ is the average absorption length. The combination $L/\lambda$ can be got by experimental measurement. There is also a more complex formula to describe the relationship between aSEY and $\theta$ \cite{Furman2002}:

\begin{equation}
	\label{equ:average_SEY_theta_relationship_furman}
	\begin{aligned}
		&\hat{\delta}\left(\theta_{0}\right)=\hat{\delta}_{t s}\left[1+t_{1}\left(1-\cos ^{t_{2}} \theta_{0}\right)\right] \\
		&\hat{E}\left(\theta_{0}\right)=\hat{E}_{t s}\left[1+t_{3}\left(1-\cos ^{t_{4}} \theta_{0}\right)\right]
	\end{aligned}
\end{equation}

Where $t_1$, $t_2$, $t_3$ and $t_4$ are all the fit parameters. The $\hat{\delta}_{t s}$ and $\hat{E}_{t s}$ are aSEY and incident energy of the SEY curve extreme point when incident electron hit on surface with an angle $0^{\circ}$. The ~\ref{equ:average_SEY} and ~\ref{equ:average_SEY_theta_relationship_furman} are all effective for description the hit angle influence aSEY value. Average SEY with arbitrary incident angle (${\rm aSEY(\theta)}$) can be obtained by the product of the average SEY when electron incident vertically (${\rm aSEY(0)}$) and an coefficient varying with $\theta$. ${\rm aSEY(0)}$ depends on the energy of incident electron and is measured experimentally for three kinds of secondary electrons.

For a single emission event, such three discrete processes are all possible to happen but they are nutually exclusive so that we should decide which process to happen by the probabilities. Since SEY must be 1 for backscattering and rediffused electron emission, the probability for these two processes in a single event will equal to ${\rm aSEY_B}$ or ${\rm aSEY_R}$. That is,
\begin{equation}
	\label{equ:probability}
	P_{\rm B} = {\rm aSEY_B},~P_{\rm R} = {\rm aSEY_R},
\end{equation}
where $P_{\rm B}$ and $P_{\rm R}$ denote the probability for backscattering and rediffused electron emission. Then the probability for true secondary emmision $P_{\rm T}$ will be:
\begin{equation}
	\label{equ:prob_true}
	P_{\rm T} = 1-P_{\rm B}-P_{\rm R},
\end{equation}

\begin{figure}[htbp]
	\centering
	\includegraphics[width=0.5\textwidth]{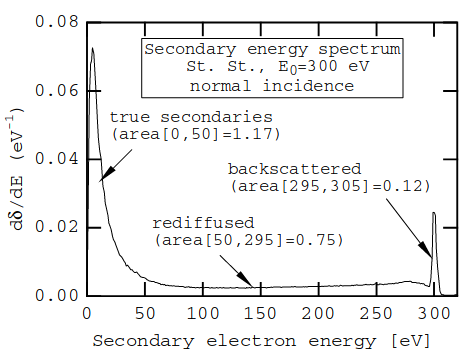}
	\caption{A sample of the measured energy spectrum.\cite{Furman2002}}
	\label{fig:distinguish_process}
\end{figure}

When characterizing the Secondary Electron Yield (SEY) at different incident electron energies, it is often challenging to explicitly distinguish the contributions from the three distinct secondary electron processes. However, a common approach is to analyze the emission energy spectrum, as illustrated in Fig.~\ref{fig:distinguish_process}. The emission energy spectrum can be divided into three regions:

\begin{enumerate}
	\item Energies below 50 eV primarily correspond to true secondary electrons.
	\item The energy range from 50 eV to 295 eV is associated with rediffused electrons.
	\item Backscattered electrons gather in the range of 295 eV to 305 eV.
\end{enumerate}

Notably, when the primary electron energy is 300 eV, the energy of backscattered electrons cannot exceed 300 eV. Any backscattered electrons with energies above 300 eV are considered to be a result of measurement error. We define the yield of backscattered electrons as $\delta_{e}$ and the yield of rediffused electrons as $\delta_{r}$. Based on the information provided in Fig.~\ref{fig:distinguish_process}, we can deduce that $\delta_{r} = 0.75$ and $\delta_{e} = 0.12$ for normal incidence with a primary electron energy of 300 eV. 

There are some empirical formulas that can approximately describe three component secondary electron yield. 

For backscattered yield $\delta_{e}$ at normal incidence\cite{bruining2016physics,redhead1968physical}.
\begin{equation}
	\delta_{e}\left(E_{0}, 0\right)=P_{1, e}(\infty)+\left[\hat{P}_{1, e}-P_{1, e}(\infty)\right] e^{-\left(\left|E_{0}-\hat{E}_{e}\right| / W\right)^{p} / p}
	\label{equ:scatter_yield}
\end{equation}

In the given equation, $P_{1, e}(\infty)$, $\hat{P}_{1, e}$, $\hat{E}_{e}$, $W$, and $P$ are fitting parameters. $E_0$ denotes electron impact energy. Equation ~\ref{equ:scatter_yield} reaches its peak at $E_0 = \hat{E}_e$. As $E_0$ tends towards infinity, the value of $\delta_e$ approaches $P_{1, e}(\infty)$.

The function of rediffused yield at normal incidence\cite{bruining2016physics,redhead1968physical}.
\begin{equation}
	\delta_{r}\left(E_{0}, 0\right)=P_{1, r}(\infty)\left[1-e^{-\left(E_{0} / E_{r}\right)^{r}}\right]
	\label{equ:rediff}
\end{equation}
Where $P_{1, r}(\infty)$, $E_r$ and $r$ are fitting parameters. The $E_0$ is the impact energy of primary electron. As $E_0$ tends to infinite, the value of $\delta_r$ approaches $P_{1, r}(\infty)$.

True secondary electron can be denoted by $\delta_{ts}$, which is well fitted by Eq. ~\ref{equ:SEY_D_function}\cite{bruining2016physics,dekker1958secondary,seiler1983secondary}.

\begin{equation}
	\delta_{t s}\left(E_0, \theta_0\right)=\hat{\delta}\left(\theta_0\right) D\left[E_0 / \hat{E}\left(\theta_0\right)\right]
	\label{equ:SEY_D_function}
\end{equation}

And $D(x)$ can be expressed: 

\begin{equation}
	D(x)=\frac{s x}{s-1+x^s}
	\label{equ:SEY_D}
\end{equation}

\begin{figure}[htbp]
	\centering
	\includegraphics[width=0.5\textwidth]{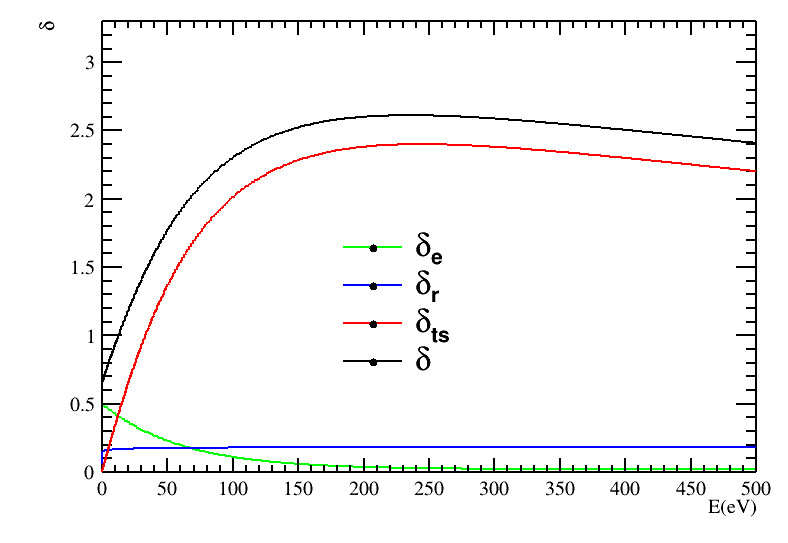}
	\caption{Curve of secondary electron production yield versus incidence electron energy for three components.\cite{Furman2002}}
	\label{fig:SEY_curve}
\end{figure}

In the Eq.~\ref{equ:SEY_D_function}, the true secondary electron yield $\delta_{ts}$ reaches its peak value $\hat{\delta}$ when the incident electron energy $E_0$ equals $\hat{E}$. Therefore, $D(1) = 1$ and $D(x)' = 0$. The variable $s$ is a fitting parameter that domain $D(x)$ function shape. Above three components SEY curve are plotted in Fig.~\ref{fig:SEY_curve}.

Then we will decide which process to happen by sampling the probabilities. The SEY in this single event will be 1 for backscattering and rediffused electron emission. For true secondary electron emission, the yield ${\rm SEY_T}$ will be got by sample a poisson distribution or a binomial distribution\cite{Furman2002} with its mean value fixed at ${\rm aSEY_T}/P_{\rm T}$ to garantee the ${\rm aSEY_T}$ is consistent for all the emission events. If SEY of a specific event is sampled to be 0, there will be no secondary electrons emitted and the incident electron will travel across the surface and interact with the material by other processes, such as bremsstrahlung, nuclear capture and so on.

\begin{figure}[htbp]
	\centering
	\subfigure[Total spectrum of secondary electrons.]{ \includegraphics[width=0.4\textwidth]{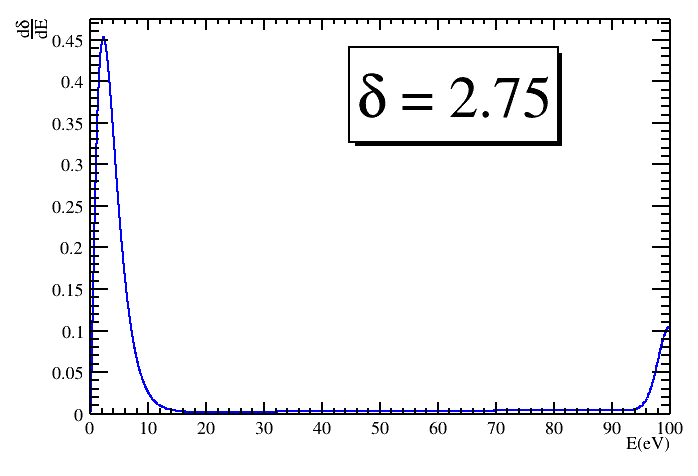} }
	\subfigure[Spectrum of backscattered secondary electrons.]{ \includegraphics[width=0.4\textwidth]{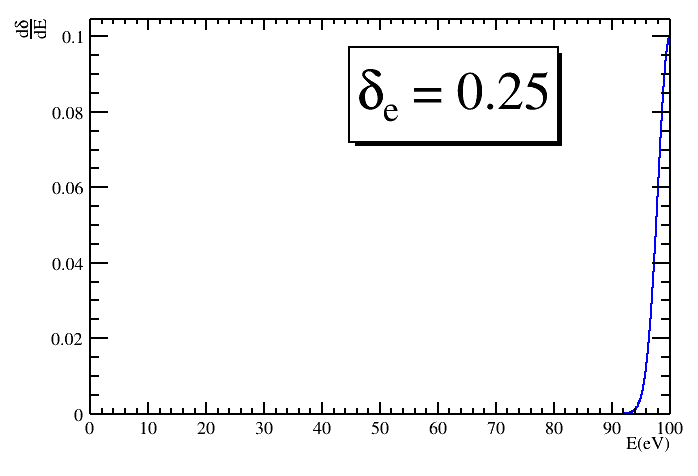} }
	\subfigure[Spectrum of rediffused secondary electrons.]{ \includegraphics[width=0.4\textwidth]{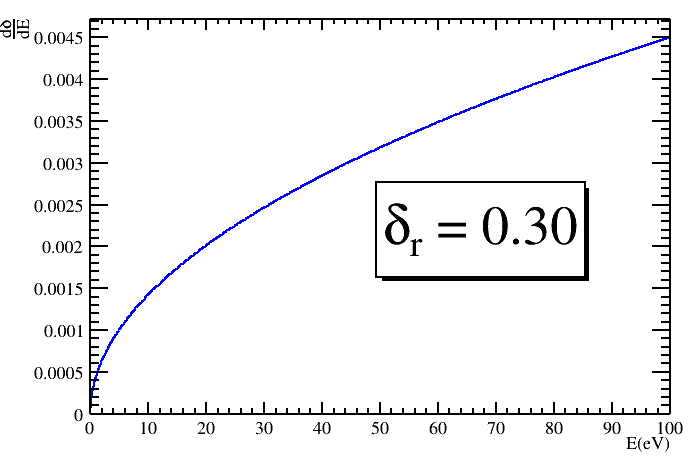} }
	\subfigure[Spectrum of true secondary electrons.]{ \includegraphics[width=0.4\textwidth]{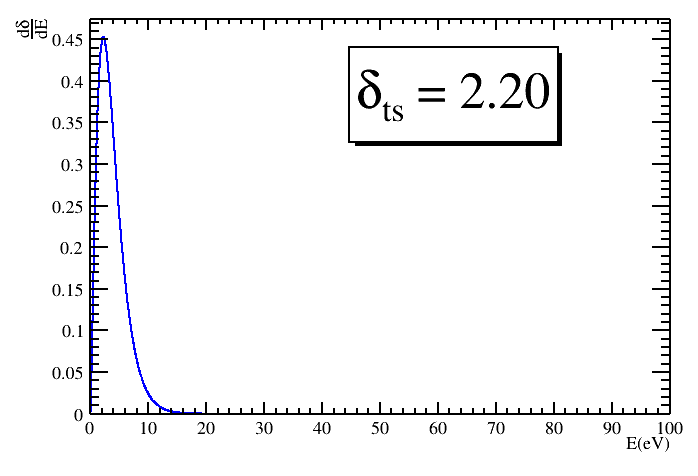} }
	\caption{Empirical formula depicts secondary electron emission spectrum for different processes.}
	\label{fig:spectra}
\end{figure}

After SEY determined, the energy of emitted electrons are obtained by sample the emission spectra of the three processes. According to Furman model~\cite{Furman2002}, the total spectrum and spectra for three specific processes follow the distributions shown as~\ref{fig:spectra}.

The spectrum of backscattered secondary electrons is described as the extend form of the Gaussain distribution with the mean value at the incident energy as shown in Eq.~(\ref{equ:backscatter_spec})
\begin{equation}
	\label{equ:backscatter_spec}
	f(E) = B\frac{2{\rm exp}[\frac{(E-E_{p})^2}{2\sigma_{e}^{2}}]}{\sigma_e \cdot {\rm erf}(\frac{E_p}{\sqrt{2}\sigma_e})},
\end{equation}
where $B$ is the normalization factor, $E_{p}$ is the incident energy and $\sigma_e$ is the standard deviation of emission spectrum got by fit to the experimental measurement. For rediffused secondary electrons, the spectrum is described as
\begin{equation}
	\label{equ:rediffused_spec}
	f(E) = C\frac{(q+1)E^{q}}{E_{p}^{q+1}},
\end{equation}
where $C$ denotes the normalization factor, $q \in (0,1)$ is a coefficient got by fitting and $E_p$ is the incident energy. As for the true secondary electrons, the spectrum is
\begin{equation}
	\label{equ:true_spec}
	f(E) = D \cdot E^{\rho_n-1}e^{\frac{-E}{\kappa_n}},
\end{equation}
where $\rho_n$ and $\kappa_n$ are both coefficients got by fitting and $D$ is the normalization factor. The spectrum of true secondary electrons does not depend on the energy of incident particle.

Since the energy must conserve during the multipication, we require the sum of all the sampled energy of secondary electrons $E_{\rm total}$ to be less than the energy of incident electron $E_p$. Once $E_{\rm total} > E_p$, the energy of secondary electrons will be resampled.

The direction of the emitted secondary electrons are obtained by sampling the angular distribution, which is the same for three kinds of secondary electrons and are expressed as
\begin{equation}
	\label{equ:angular_distribution}
	\frac{{\rm d}P}{{\rm d}\phi} = A e^{-\frac{c}{{\rm cos}\phi}},
\end{equation}
where $P$ denotes the possibility of secondary electron emission, $\phi \in [0,\frac{\pi}{2})$ is the angle between emitted particle and the normal direction of the surface, $c$ is a coefficient got by experiment and $A$ denotes the normalization factor.

The essential parameters of emitted secondary electrons will be transferred to the Geant4 kernel and generate new particles as described, after which the incident electron will be killed by our process.

\begin{figure}[htbp]
	\centering
	\subfigure[SEY of true secondary electron emission with respect to $\theta$ angle.]{ \includegraphics[width=0.45\textwidth]{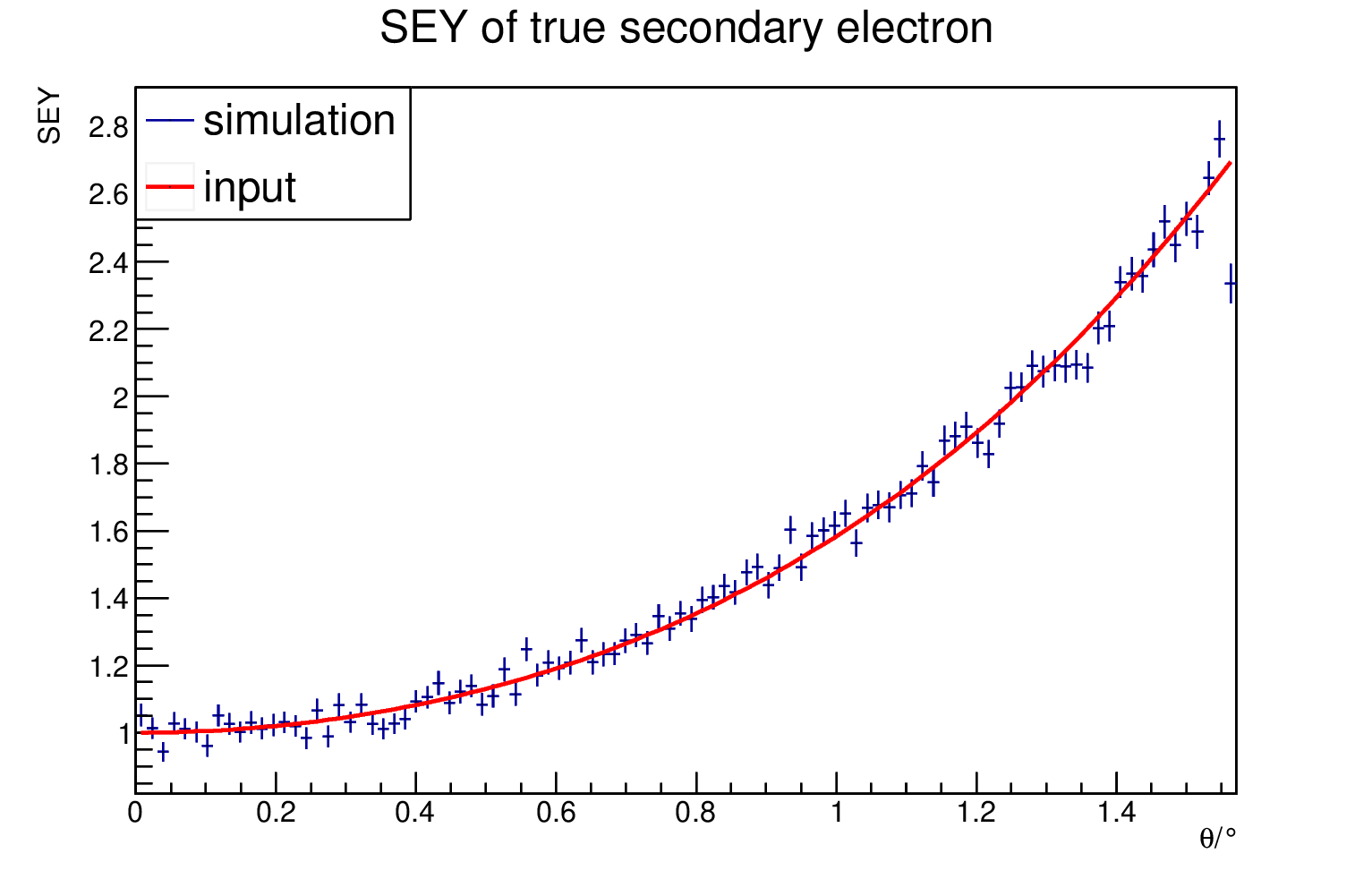} }
	\subfigure[SEY of true secondary electron emission with respect to the energy of incident electron.]{ \includegraphics[width=0.45\textwidth]{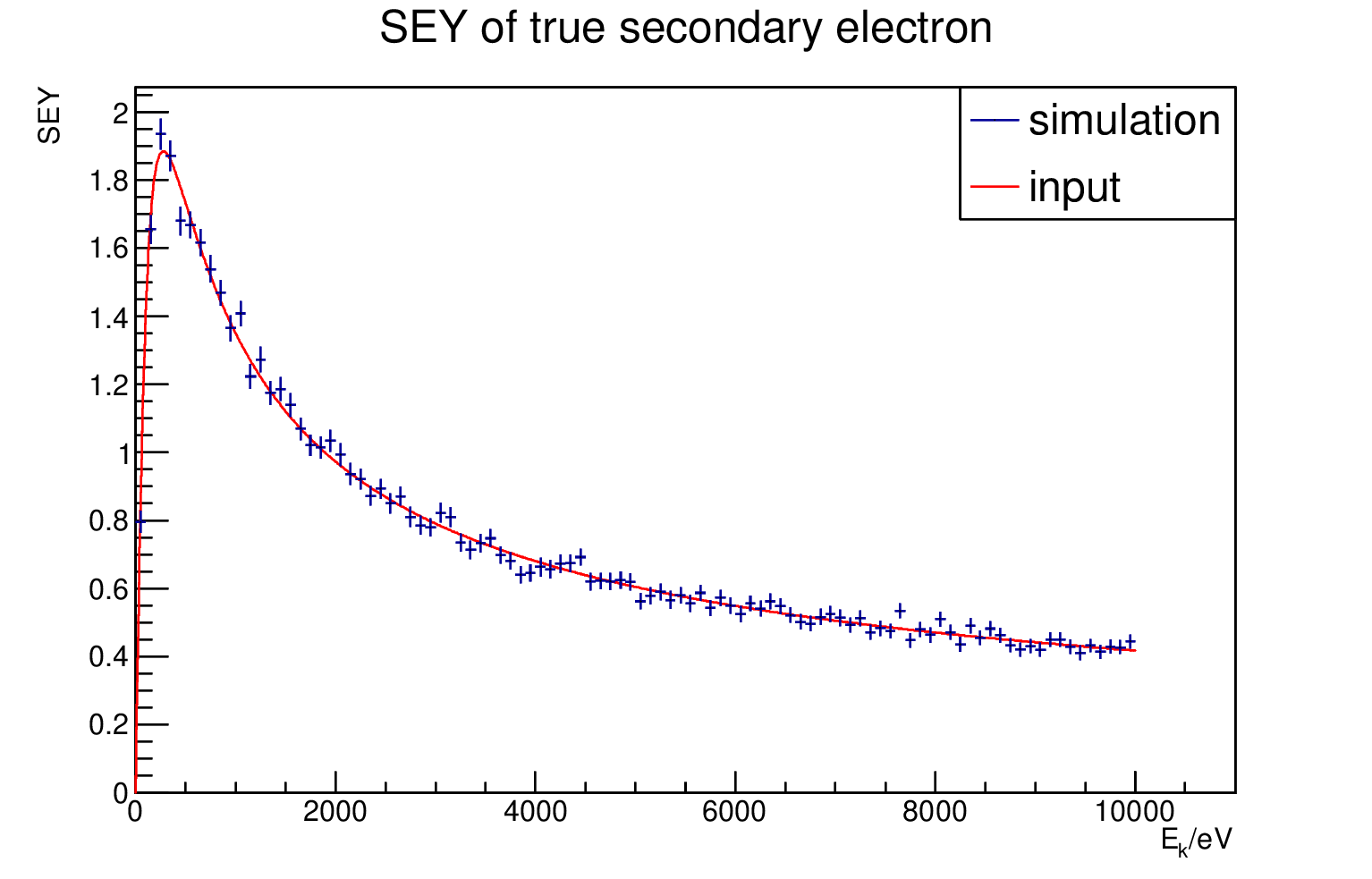} }
	\label{fig:SEY_true}
	\caption{Comparison of SEY distribution of secondary electron.}
\end{figure}

To varify our generator in Geant4, we compared multiple distributions of various parameters between the simulation and the input value. Since the true secondary electron emission is the most important part that contribute most of the secondary electrons during the multiplication, we present the SEY distribution of true secondary electrons in Fig.~\ref{fig:SEY_true} with respect to $\theta$ angle mentioned above and the energy of incident electron. The comparison shows excellent consistency between input parameters and the simulation results.

The descriptions above are the case where an electron hit the surface of material, which is the most common situation in a cascade multiplication of electrons. In the case where other kinds of particles incident, the secondary electron emission process will still work if there are electrons with energy within the multiplication range, usually $[1, 10000]~{\rm eV}$, are created during the interactions processed by Geant4, such as photoelectric effect, Compton scattering and so on.

To fasten the secondary electron emission process in the real calculation, some correponding mathematics functions of Boost C++ library~\cite{Boost} are utilized in MCPSim.

\section{Configuration of simulation}
\label{sec:geo_and_field}

\subsection{Geometry}
MCPSim provide multiple methods for users to define the geometry of their devices. The following parts is a brief introduction. The detail information can be found in the user manual.

\subsubsection{Define the geometry by code}
Since MCPSim is an open-source toolkit, users are naturally allowed to define the geometry of their devices by directly modify the code. The method is the same as other Geant4-based simulation program so that users can just refer to the user manual of Geant4~\cite{Geant4}. It has to be mentioned that users are responsible to define the sensitive detectors by themselves, or there will be no output information after simulation.

\begin{figure}[htbp]
	\centering
	\includegraphics[width=0.5\textwidth]{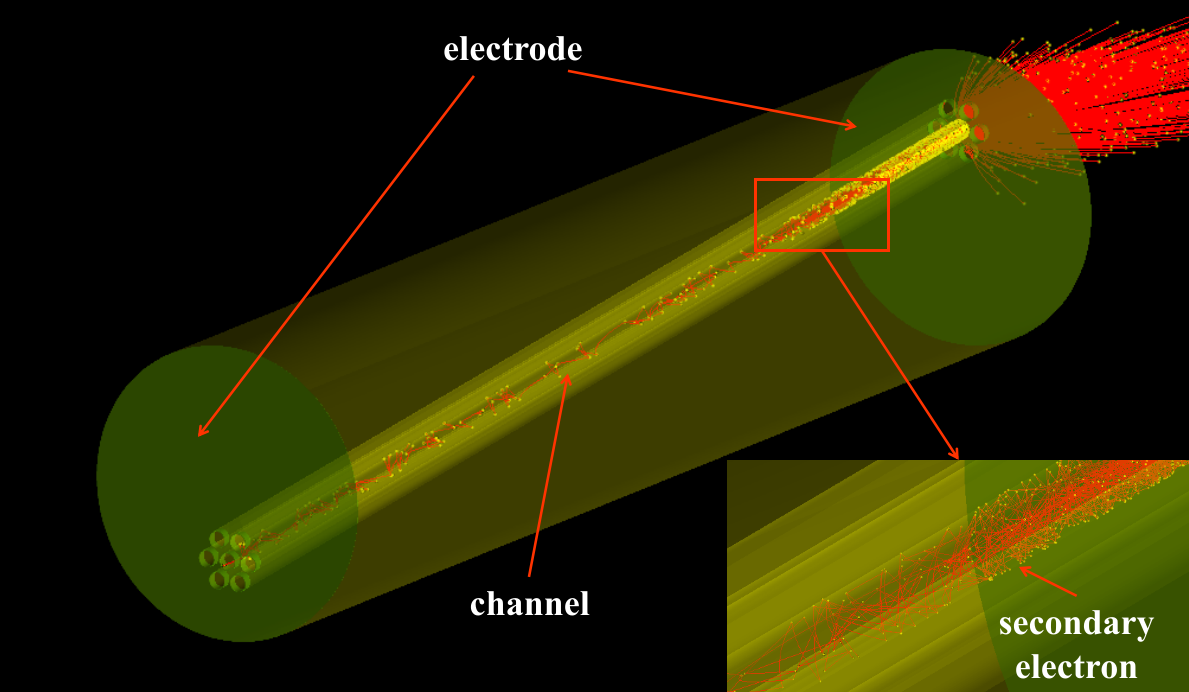}
	\caption{Schematic picture of the simulation using simple geometry defined directly by code.}
	\label{fig:simulation_pic}
\end{figure}

Fig.~\ref{fig:simulation_pic} shows a simple MCP geometry including the electrodes of both sides and the channels defined by code, which is much more trivial than a real MCP but is enough and precise for ones to study the response of a single or several channels that is usually the key concern in the optimazation of MCP. In Fig.~\ref{fig:simulation_pic}, the field is set to be uniform and the incident electron is multiplied within the channel.

\subsubsection{Define the geometry by necessary parameters specially for MCP}

For the special case of MCP, we provide the interface for users to define a general geometry by some necessary parameters so that ones are able to configure the simulation fast and easily. The essential parameters are shown in Tab.~\ref{tab:para_geom}. Each parameter listed in the following table has a default value to garantee MCPSim to work normally. However, you may get wrong results if you use the default value. More detailed information about the configuration can be found in our user manual.

\begin{table}[htbp]
	\centering
	\doublespacing
	\caption{Essential parameters for building MCP geometry.}
	\begin{tabular}{cc}
		\hline
		\hline
		Parameter	&	Description	\\
		\hline
		\texttt{MCP\_radius}	&	The total radius of the MCP	\\
		\texttt{MCP\_thickness}	&	The thickness of the MCP	\\
		\texttt{channel\_radius}	&	The radius of the channels	\\
		\texttt{channel\_rotate}	&	The angle of inclination of channels	\\
		\texttt{electrode\_thickness}	&	The thickness of the electrode	\\
		\texttt{distance\_channel}	&	The distance between the center axises of two neighbour channels	\\
		\texttt{other\ options}	&	\\
		\hline
		\hline
	\end{tabular}
	\label{tab:para_geom}
\end{table}

\subsubsection{Import geometry from CAD files}
\begin{figure}[htbp]
	\centering
	\includegraphics[width=0.75\textwidth]{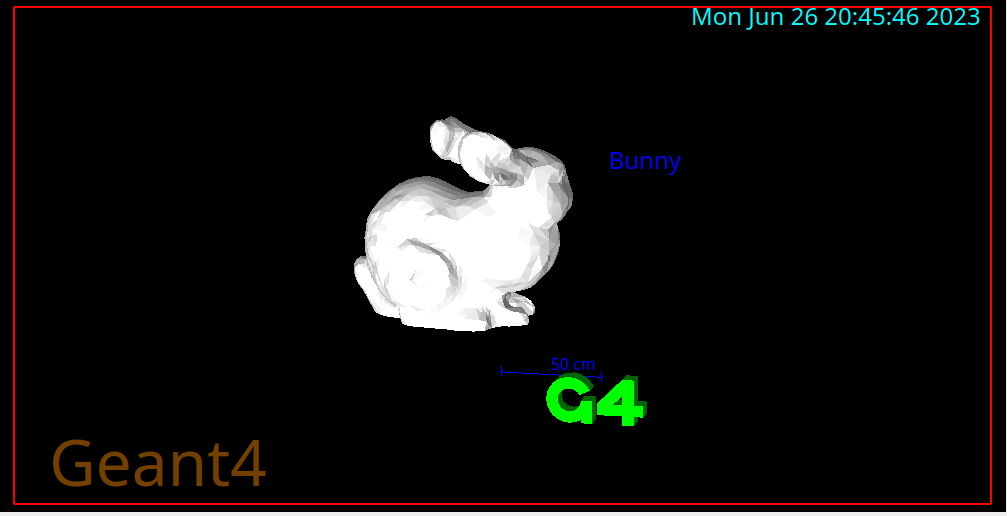}
	\caption{The example for importing a complex geometry into the simulation.}
	\label{fig:CADimport}
\end{figure}

As mentioned above, defining geometry by directly coding requires fundamental knowledge of programming and Geant4. The code will be much complex and hard to handle if the geometry is not a combination of regular geometries like box, cylinder etc. Meanwhile, the second method to define only essential parameters from the user interface can only be used when you want to simulate MCP. Both methods have obvious shortages for practical applications. To solve this problem, we combined the toolkit CADMesh~\cite{Poole2012} to MCPSim to import CAD files to the simulation. Fig.~\ref{fig:CADimport} shows a simple sample. It should be noticed that only solids can be imported into Geant4, while the combination relationship and the material of the solids are still the responsibility of users. Detailed instruction for this method can be found in the user manual.

\subsection{Field}

The simulation of electron multipliers will be influenced a lot by the precision of electromagnetic fields and the tracking of electrons in fields. MCPSim provide three method for users to define the electromagnetic fields in the simulation. For the definition of electromagnetic fields, MCPSim provides three methods for users. The tracking of electrons in fields can also be tuned by adjusting the steppers or parameters provided by Geant4, which is wrapped in the user interface of MCPSim introduced in our user manual.

\subsubsection{Uniform field for MCP}

Actually, for a MCP whose length-to-diameter ratio is not so small, the electric field inside the channels are very close to a uniform field. Therefore, we provide an interface for users to conveniently define a uniform electric field of MCP by the high voltage between electrodes or the value of field strength. Users can also define a uniform magnetic field by the same way. This method is valid only for MCP simulation.

\subsubsection{Electric field calculated numerically for MCP}

If users need to simulate the behavior of electron multiplication in MCP more precisely, we also developed a component to calculate the field with given high voltage using the Finite Element Method(FEM). Users only need to define the voltage before initiating the simulation. However, this approach imposes a significant computational burden on tracking secondary electrons. To facilitate the simulation process, FEM is employed, starting with the meshing of the relevant spatial region. A suitable open-source mesh generator, such as gmsh, can be utilized for this purpose. gmsh offers numerous C++ APIs that can be embedded into our MCPSim software. By using the delaunay triangulation algorithm in gmsh, the geometry can be meshed, and an msh format file containing essential node relationship information can be exported.

For this study, a generalized poisson \cite{nagel2011solving} need to be solved using FEM. While the finite difference method is generally preferable for solving the potential within a single material domain with a dielectric constant. However, in MCP simulation, the high voltage only applies to two plane surfaces. As a result, potential calculations are required in both the MCP body ($\epsilon_1$) and the channel ($\epsilon_2$). The interface between the MCP body material ($\epsilon_1$) and the channel material ($\epsilon_2$) introduces a discontinuity in the secondary derivative of the Poisson equation. FEM does't calculate Eq.~\ref{equ:possion} directly. Generally, FEM employs a conversion of the Poisson equation to its weak form \cite{ottosenintroduction}, which effectively reduces it to a first-order equation.

In FEM, the final step involves solving a large matrix equation of the form $\mathbf{K}\mathbf{x} = \mathbf{f}$. In this equation, $\mathbf{K}$ is called stiffness matrix and $\mathbf{f}$ is called load matrix, $\mathbf{x}$ is unknown variable. Stiffness matrix is typically a  sparse matrix. We utilize Eigen3 library in which is contained many iterative solution algorithm to solve the matrix equation.

\begin{equation}
	\label{equ:possion}
	\nabla \left(\epsilon \nabla \phi \right) = 0
\end{equation}

\begin{figure}[htbp]
	\centering
	\subfigure[The model to test our code FEM performance.]{ 
	\includegraphics[scale = 0.5]{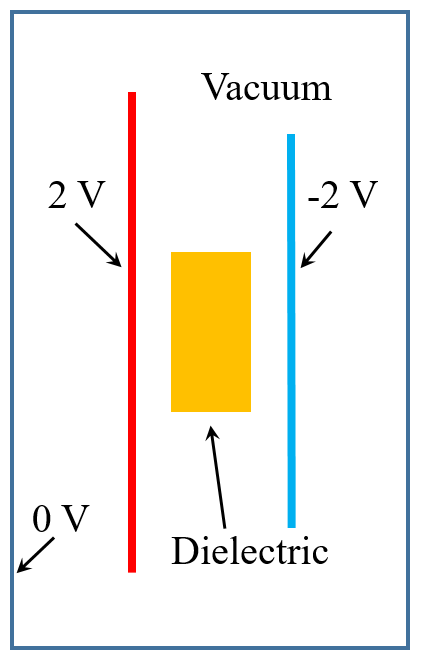} }
	\subfigure[Delaunay triangulation of test model.]{ 
	\includegraphics[scale = 0.5]{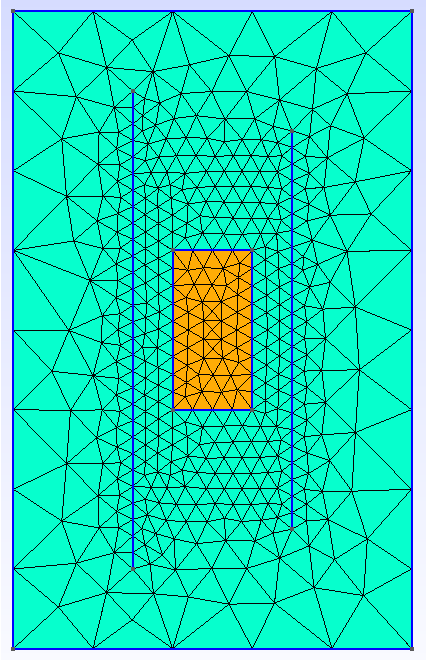} }
	\label{fig:benchmark}
	\caption{The benchmark of FEM to calculate electric field.}
\end{figure}

\begin{figure}[htbp]
	\centering
	\subfigure[Our FEM calculation solution.]{ \includegraphics[width=0.45\textwidth]{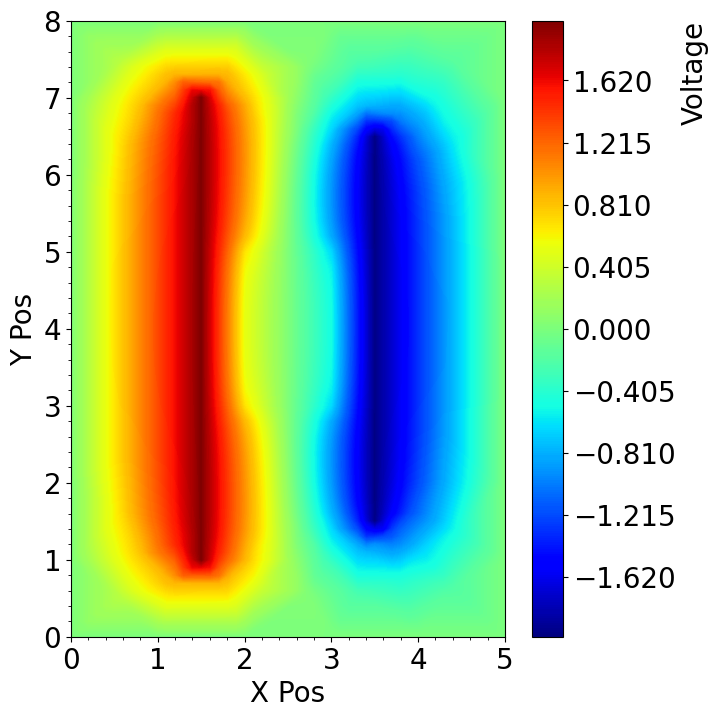} }
	\subfigure[Comparison with commercial software solution.]{ \includegraphics[width=0.45\textwidth]{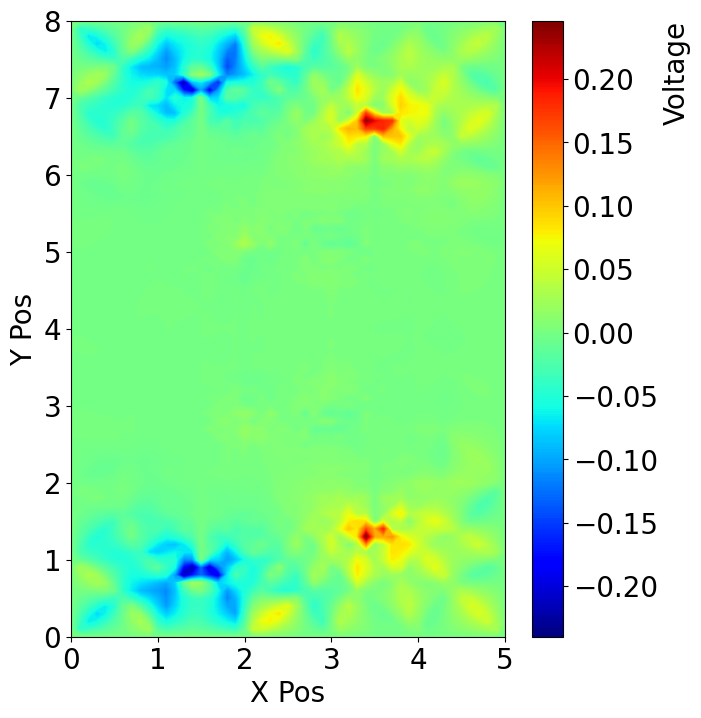} }
	\label{fig:error}
	\caption{Solution.}
\end{figure}

Fig.~\ref{fig:benchmark} shows a benchmark model. It consist of two parallel plane, one is apply 2V and another apply -2V. A dielectric material is placed between these planes to examine the interface performance with varying $\epsilon$ values. The potential distribution in both the vacuum and dielectric regions is calculated using the FEM. The obtained results are then compared to those from a commercial FEM software, which is considered the standard solution. Analyzing the residual error presented in Fig.~\ref{fig:error}, it is evident that a significant error occurs at the vertices of the model, while the error surrounding dielectric is relatively small. This discrepancy can be attributed to the different grid generation methods and shape functions employed in FEM calculation process.

\subsubsection{Implement by a field map}

\begin{figure}[htbp]
	\centering
	\doublespacing
	\includegraphics[width=0.7\textwidth]{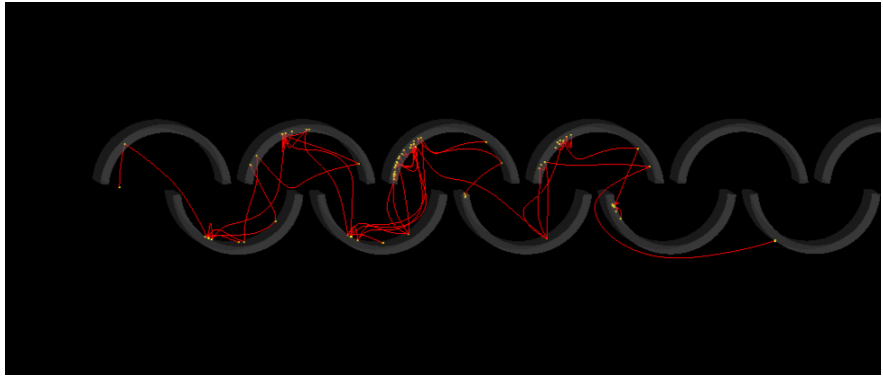}
	\caption{The schematic picture of the simulation of an electron multiplier with several dynodes, where the electric field is applied by the imported field map. The number of tracks is limited to several tens in this picture.}
	\label{fig:dynode_sim}
\end{figure}

For devices except from MCP, we provide the interface for users to define the field by a field map. Since the structure of various devices may be different a lot, the field should be calculated by users and written in a given format to import to MCPSim. The field map must consist of the field points (vertice of box-shaped elements divided uniformly inside the whole field) and the corresponding strength of each point. The shape of the whole field should be a box. The standard file of field map in MCPSim is a ROOT file including two trees. One is the necessary information of the field, such as the size of the field elements, the number of the elements along the xyz axis and the start point of the field, and the other tree includes the positions of field points and the corresponding field strength. The method has been tested successfully to a model of an electron multiplier with several dynodes as is shown in Fig.~\ref{fig:dynode_sim}.

\subsection{Particle generator}

We provide users two methods to define the particle source, which are realized by directly applying the \texttt{G4ParticleGun} and \texttt{G4GeneralParticleSource (GPS)} in Geant4~\cite{Geant4}. Users can define a simple monochromatic source without specialized beam structure by the method inherited from \texttt{G4ParticleGun}. The second method inherited from \texttt{G4GeneralParticleSource} allows users to define a much more complex particle source complemented with the shape, the angular distribution and the energy spectrum of the source. Multiple sources are also able to be added to a single simulation. The detailed instructions can be found in our user manual.

\subsection{Parameters for secondary electron emission process}

As mentioned in Section~\ref{sec:Furman}, Furman model depends a lot on experimental measurement so that there is a large set of parameters necessary to be determined. In MCPSim, we provide the parameters for several typical materials. 
If users need to adjust or optimize the secondary emission model, interfaces are provided to change the parameters mentioned in Section~\ref{sec:Furman}, including $L/\lambda$ for the dependence of SEY with respect to the incident angle, $\sigma_e$ for spectrum of backscattered secondary electrons, $q$ for rediffused electrons, $\rho_n$ and $\kappa_n$ for true secondary electrons, $c$ for the angular distribution of emitted electrons.

\subsection{Other options}
Except the options and parameters mentioned above, MCPSim also provides interface for users. For example, users are able to control the output level of the information, warning and errors during the simulation by the parameter \texttt{verbose\_level}. The output data structure introduced in Section~\ref{sec:event_model} can also be chosen by users. Ones are also able to limit the maximum number of tracks to be simulated.

\subsection{User interface}
The user interface of MCPSim is inherited from Geant4~\cite{Geant4}. All the settings and parameters mentioned above can be modified using the interfaces. Users are allowed to configure the simulation by either a macro file or the commands in the UI of Qt5 implemented by Geant4. The UI window is convenient for ones to adjust the simulation in order to get the correct results. Every track will be presented visibly like Fig.~\ref{fig:simulation_pic} but the speed of simulation will be significantly slowed down due to the graphic rendering. Therefore, a limit of number of tracks is strongly suggest when UI window is used. If users need only to change several simple parameters, it may be much easier to directly use the UI method. If complex configuration are necessary or users would like to submit works to computer clusters, using a macro file is advised. Detailed instructions can be found in our user manual.

%

\section{Output data structure}
\label{sec:event_model}

The output file is designed to be a ROOT~\cite{ROOT} file containing the essential information for users to preform their analysis. The accessible information is listed in Tab.~\ref{tab:event_structure}. Users could follow the user manual 
to handle the output file.

Except the ROOT file format, users can also choose to output the information in a plain text (.txt) file but it is not suggested since it will be difficult to handle and very long when there are large quantities of secondary particles. 

\begin{table}[htbp]
	\centering
	\doublespacing
	\caption{Accessible information in the output file of MCPSim.}
	\begin{tabular}{cc}
		\hline
		\hline
		Variable			&		Description		\\
		\hline
		\texttt{EventID}		&	The index of events		\\
		\texttt{TrackID}		&	The index of particle tracks	\\
		\texttt{Pdg}			&	PDG code~\cite{osti_5545302, Knowles:1995kj, ParticleDataGroup:1998hll} of the particles	\\
		\texttt{CreatorProcess}		&	The process creating the particle	\\
		\texttt{initP}			&	The initial momentum of the particle	\\
		\texttt{finalP}			&	The final momentum of the particle	\\
		\texttt{initPos}		&	The initial position of the particle	\\
		\texttt{finalPos}		&	The final position of the particle	\\
		\texttt{detP}			&	The momentum of particles entering detector	\\
		\texttt{detPos}			&	The position where particle hit the detector	\\
		\hline
		\hline
	\end{tabular}
	\label{tab:event_structure}
\end{table}

\section{Secondary electron low energy area yield}

\begin{figure}[htbp]
	\centering
	\doublespacing
	\includegraphics[width=0.5\textwidth]{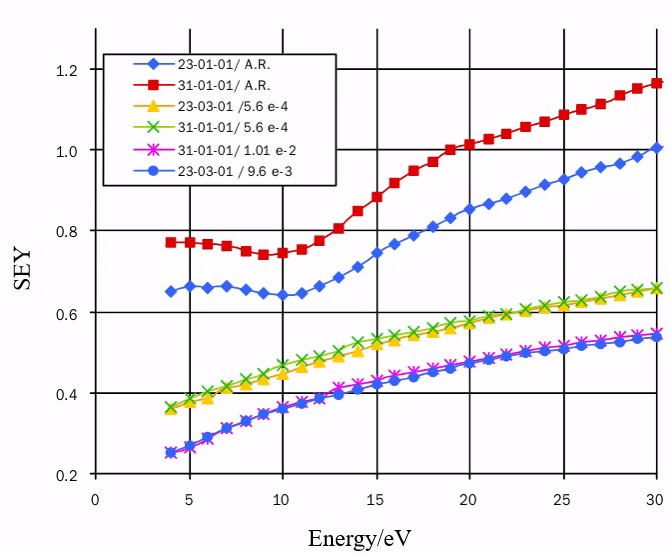}
	\caption{The average secondary electron yield versus electron energy for copper samples when the incident electron energy lower than 30 eV\cite{baglin2001summary}.}
	\label{fig:LowEnergy}
\end{figure}

For secondary electron low energy range (primary electron energy lower than 30 eV) yield, it is a challenges to measure accurately in our laboratory. However, in certain case, the simulate result is strong sensitivity on low energy details of the SEY and energy spectrum \cite{Furman2002}. To achieve more realistic simulation results, it is crucial to understand the behavior of secondary electrons in the low energy range. The Furman model utilizes Eq.~\ref{equ:scatter_yield}, Eq.~\ref{equ:rediff} and Eq.~\ref{equ:SEY_D_function} to describe three components of secondary electron yield, and the total SEY is the sum of these three functions. By applying these experimental formulas and fitting them to test data, we can extrapolate the SEY curve to the low energy range.
However, the accuracy of this extrapolation can only be verified through experimental data. In fact, SEY curve data for copper material in the low energy range has been measured and is presented in Fig.~\ref{fig:LowEnergy}. The data suggests that the SEY curve tends to a constant value when impact energy below 10 eV, indicating a discrepancy between the extrapolated data and the test data.

From the perspective of true secondary process, it can be decomposed three steps. Firstly, a primary electron with a certain energy bombards material surface and penetrates to a depth of about nanoscale length\cite{zhu2020theoretical}. Next, the primary electron deposit energy along its path. causing some electrons in the material absorb energy and convert to excited state. A portion of excited electron spread to the surface and overcome surface potential barrier. Finally, these electron emit from the material become true secondary electron. Based on these steps, we can find that the excited electron must have sufficient energy that is greater than material work function. Consequently, being similar with photoelectric effect, only if the primary electron energy exceeds a certain value, true secondary electron emits from material. Therefore the $\delta_{ts}$ curve  needs to be modified accordingly.

\begin{figure}[htbp]
	\centering
	\doublespacing
	\includegraphics[width=0.5\textwidth]{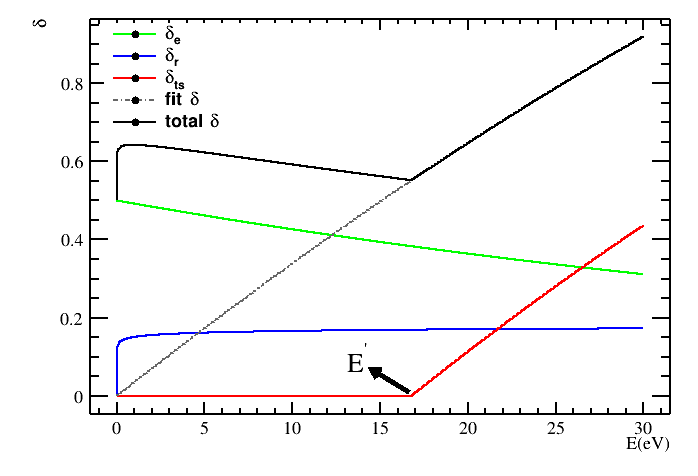}
	\caption{True secondary electron yield curve.}
	\label{fig:LowEnergySEYCurve}
\end{figure}

According to goodness of fit of Eq.~\ref{equ:SEY_D}, we can find that the formula describes the total SEY curve very well in the high energy range (>100 eV). Focusing on $\delta_{r}$ and $\delta_{e}$ curves in Fig.~\ref{fig:SEY_curve}, they tend to converge to the constant values of  $P_{1,e}(\infty)$ and $P_{1,r}(\infty)$ respectively in the high energy range. Hence, the true secondary yield cure can be obtained by shifting the total SEY curve downward by  $P_{1,e}(\infty)+P_{1,r}(\infty)$. As a result, we will get a curve that intersects the x-axis at the energy $E^{\prime}$, as shown in Fig.~\ref{fig:LowEnergySEYCurve}. Reflecting on the process of true secondary electron generation, it can be inferred that the $E^{\prime}$ represents the minimum incident electron energy required to generate true secondary electrons. When incidence electron energy less than $E^{\prime}$, $\delta_{ts} = 0$.

\section{Comparison with experiment results}
\label{sec:comparison}

\begin{figure}[htbp]
	\centering
	\doublespacing
	\includegraphics[width=0.5\textwidth]{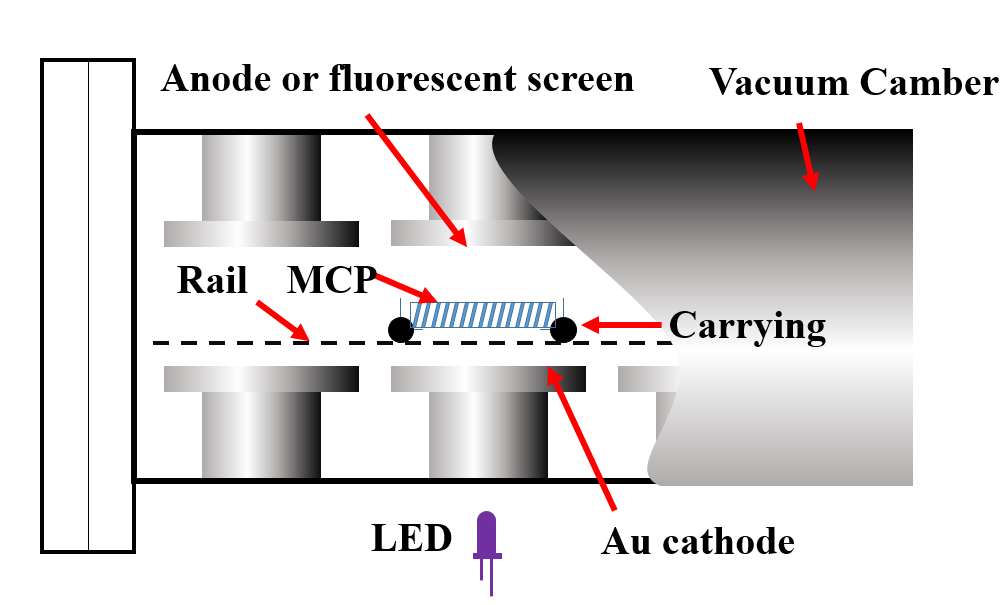}
	\caption{Schematic of MCP testing device.}
	\label{fig:device}
\end{figure}

In laboratory, a vacuum device 
was used to measure the gain of MCP  at different applied voltages. Its diagram is shown in Fig.~\ref{fig:device}. A low current in the order of pA is emitted from an electron source located at the bottom of the chamber. The degree of uniformity of MCP or the output current is observed either on a fluorescent screen or measured using a galvanometer. In this paper, MCPs with different geometrical parameters were measured, and the specific parameters are listed in Tab.~\ref{tab:mcp_par}. For \# 1 MCP, its body resistance is 84 M$\Omega$. During the test, an input current of 107 pA was applied, and the operating voltage was varied from 700V to 1200V. Similarly, for \# 2 MCP, body resistance is 90 M$\Omega$ and test input current is 164 pA, and the operating voltage ranged from 600V to 1500V. 

\begin{equation}
	\label{equ:gain_def}
	G = \frac{I_{out}}{I_{in}}
\end{equation}

\begin{table}[htbp]
	\label{tab:mcp_par}
	\centering
	\doublespacing
	\caption{geometry parameters of MCP}
	\begin{tabular}{ccc}
		\hline
		\hline
		Geometry parameters 	& \#1 MCP 		& \#2 MCP \\
		\hline
		Thickness 				& 0.42 mm		& 0.48 mm\\
		Out diameter 			&25 mm 			&24.8 mm\\
		Pore diameter 			& 10 $\upmu$m 	& 6 $\upmu$m\\
		Bias angle 				& $10^{\circ}$ 	& $5.5^{\circ}$\\
		\hline
		\hline
	\end{tabular}
	
\end{table}

Using MCPSim, we simulated the emission of 1000 electrons from the same location with identical momentum towards the input surface of the MCP. The resulting output electrons were recorded and organized according to the data structure presented in Tab.~\ref{tab:event_structure}.By comparing these simulation results with the experimental measurements, we observed a close agreement between the two sets of data.

\begin{figure}[htbp]
	\centering
	\subfigure[MCP \#1 experimental data versus simulation data.]{ 
		\includegraphics[width=0.4\textwidth]{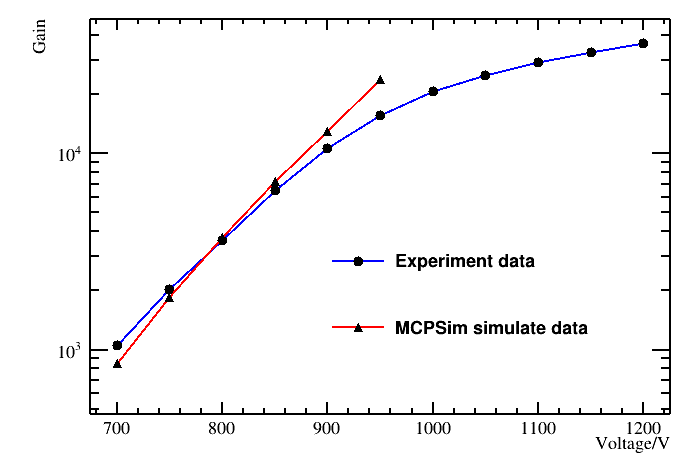} }
	\subfigure[MCP \#2 experimental data versus simulation data.]{ 
		\includegraphics[width=0.4\textwidth]{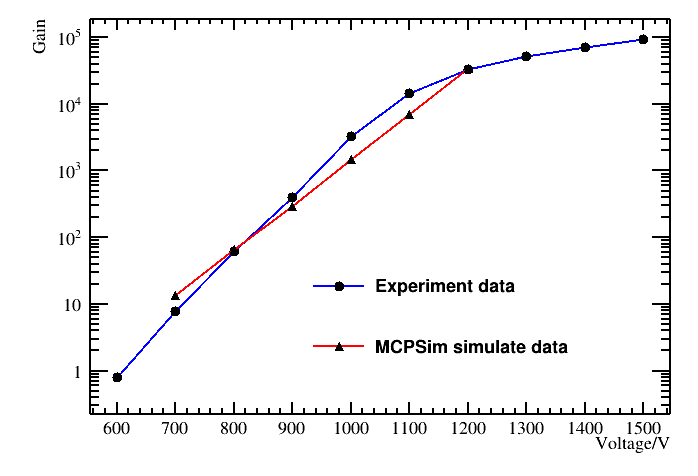} }
	\label{fig:SimulateAndMeasurement}
	\caption{MCP gain versus apply voltage curve.}
\end{figure}

The Fig.~\ref{fig:SimulateAndMeasurement} show gain versus voltage relationship for \#1 and \#2 MCPs. Upon comparing with simulation results with experimental data, we can find in the low gain range, both gain and voltage follow an exponential law. Specifically for  MCP \#1, the experimental and simulation data exhibit good consistency before reaching 900V. However beyond 900V, the gain of MCP \#1 deviates from the exponential law, and the discrepancy between the simulation and experimental data increases. This divergence can be attributed to the weakening of gain tendency caused by saturation effects.

\section{Summary}
\label{sec:summary}

In this work, a open-source, generic simulation toolkit for electron multipliers, especially for MCP, is developed based on Geant4, ROOT, Boost and other external packages. Abundant interfaces are provided for ones to configure their simulation quickly and simply, which makes MCPSim a very flexible package for anyone who need to simulate devices involved with secondary electron emission process. Good agreements are found in the comparison between simulation results and experimental measurements. In view of the exciting application of MCP and other electron multipliers in multiple fields including high energy physics, astrophysics, radiography and chemical imaging, MCPSim is expected to be a useful toolkit for the design and optimization of such devices.

In future, MCPSim will be continuously maintained and developed to satisfy the requirements of more situations. Factors influencing the electron multiplication, such as saturation, space charge and other more complicated effects, will be considered much more realistically. We hope MCPSim to bring convenience to the scientific and engineering researches.

\section*{Acknowledgements}
The authors would like to express thanks to Kaile Wen, Jian Tang and Ye Yuan for their strong support, significant effort and precious comments.

This work is supported by the National Natural Science Foundation of China (NSFC) under Contracts Nos. 11975017, 11935018, U1832204; the State Key Laboratory of Particle Detection and Electronics (SKLPDE-ZZ-202215); International Partnership Program of Chinese Academy of Sciences, Grant No. 113111KYSB20190035.





\bibliographystyle{elsarticle-num}
\bibliography{references}






\end{document}